\documentstyle[12pt]{article}
\date{ }

\title{Does noncommutative geometry predict nonlinear
Higgs mechanism?}

\author{J. S\l adkowski $^{\dag \ *}$  \\
Fakult\" at f\" ur   Physik Universit\"at Bielefeld,\\
D-4800 Bielefeld 1, Universit\"atsstrasse 25, Germany}
\begin{document}
\baselineskip9mm
\maketitle
\begin{abstract}
\baselineskip9mm

It is argued that the noncommutative geometry construction of the
standard model predicts a nonlinear symmetry breaking mechanism
rather than the orthodox Higgs mechanism.
Such models have experimentally verifiable consequences.

\end{abstract}

\ \ \ \ \ \ \ PACS number(s): 12.15.Cc
\vspace{30mm}

$^{\dag}$ A. von Humboldt Fellow; permanent address:
Institute of Physics,
University of Silesia, Pl 40007 Katowice, Poland.

$^{*}$ Internet:
sladk@usctoux1.cto.us.edu.pl and sladk@hrz.uni-bielefeld.de

\newpage
\ \ \ The unification of electromagnetic and weak interactions
is one of the biggest achievements of theoretical physics. It is
usually referred to as the Glashow-Weinberg-Salam model
(GSW model). This model successfully describes all known
experiments involving electroweak interactions.  We do
believe that the existence the Higgs particle and the missing
members of the the third family will be soon confirmed.
The situation is far less satisfactory from the theoretical
point of view: the model contains too many free parameters
and the symmetry breaking sector is put ad hoc. String theory
[1] may provide us with an explanation for the nature of
(light) generations [2-4]. Recently, new ideas have been put
forward [5-11].  They make use of the A. Connes noncommutative
geometry [12]. A Connes managed to reformulate the standard
notions of differential geometry in a pure algebraic way that
allows to get rid of the continuity and differentiability. As
there is geometrical interpretation of gauge theory in terms of
fiber bundles and connections on them, one can also apply
this formalism to the GSW model [5-9, 11] and grand unification
[10]. The notion of spacetime manifold M described by the
(commutative) algebra of functions on M can be generalized to
(a priori) an arbitrary noncommutative algebra. Fiber bundles
become projective modules. A properly generalized connection can
describe gauge field on these algebraic structures. The reader
is referred to [5-12] for details. This allows to incorporate
the Higgs field into the gauge field and the correct (leading
to spontaneous symmetry breaking) form of the scalar potential
is obtained in a natural way, provided there are at least two
generation of fermions! This sort of unification determines also
the  (classical) value of the Weinberg angle. One can add QCD
to the model in such a way that the full standard model is
reproduced. The Lagrange function one gets has the orthodox
form with the above predictions. Of course, these predictions
may get renormalized after quantization. Toy models suggest
that it is difficult to keep the relations intact. Probably
one should invent a noncommutative generalization of quantization
in order to exploit the noncommutative character of the approach.

\ \ \ Here we would like to point out that the noncommutative
generalization may predict a nonlinearly realized spontaneous
symmetry breaking, known under the acronym BESS (breaking
electroweak sector strongly) [14, 15]. Our main argument for
BESS can be stated as follows. The noncommutative version of
the standard model predicts the required form of the Higgs
sector but fermion masses (Yukawa couplings) and the number
of generation, $N_{g}$, are free parameters. There must be at
least two generations but why not, say, 127? It is natural to
suppose that $N_{g}$ is big or even unlimited and that the
fermion masses emerge as a result of interaction and the spacetime
structure. We see only the lightest fermions because the energy at
our disposal is not high enough. The Higgs particle has not yet
been discovered. Does it really exist as a physical particle? We
will show that it can be thought of in the
limit $m_{H} \rightarrow
\infty$. The main argument against BESS is that such models are
nonrenormalizable. Noncommutative geometry says that our notion
of spacetime is only an approximation (an effective electromagnetic
spacetime). The correct description is in terms of algebras.
Should we not give up the requirement of renormalizability?
BESS models can certainly lead to physical prediction [14, 15].
General relativity provide us with analogous arguments [16].

\ \ \  We will consider a noncommutative space $(A, h, D,
\Gamma )$ where  $A$ is an involutive algebra, $h$ a Hilbert space,
$D$ and unbounded self adjoint operator on $h$ (Dirac operator)
and $\Gamma$ a grading such that $A$ is even and $D$ odd. $\Gamma$
will provide us with the $\gamma _{5}$ matrix [5-12]. We shall
choose $A$ to be the algebra that corresponds to the
two-points-extension of the spacetime [5-11, 17]:

$$ A= C^{\infty}\left( M\right)\otimes A_{2}\ , \eqno(1)$$
where $C^{\infty}(M)$ is the algebra of functions on the spacetime
(spin-) manifold $M$ and $A_{2}$ is the direct sum:

$$A_{2}= {\bf H}\oplus {\bf C} \eqno(2)$$
of quaternions ${\bf H}$ and complex numbers ${\bf C}$. The Hilbert
space $h$ has the form

$$L^{2}(S(M))\otimes{\bf C}^{N_{g}} \ , \eqno(3)$$
where $L^{2}(S(M))$ denotes the Hilbert space of the
square-integrable spinors (completion of the sections of the spin
bundle $S(M)$). The total fermion space has the form:

$$L^{2}(S(M)) \otimes {\bf C^{7N_{g}}} \ . \eqno(4)$$
This corresponds to the fermions written in the form:

$$\psi = \pmatrix{\nu_{L} \cr e_{L} \cr u_{L} \cr d_{L} \cr e_{R}
\cr u_{R} \cr d_{R} \cr} = \pmatrix{\psi _{L}\cr \psi _{R} \cr }
\ , \eqno(5)$$
where each entry describes $N_{g}$ ordinary fermions. So far, we
have only considered the electroweak sector of the standard model.
The QCD part should be added "in a commutative way" because the
$SU(3)_{color}$ is an unbroken symmetry [5-8, 11]. To this end one
have to consider the algebra:

$${\bar A}=C^{\infty}\left( M\right) \otimes A_{c}=
C^{\infty}\left( M\right) \otimes A_{2}\otimes \left( {\bf C}
\oplus {\bf C}^{3\times 3}\right)\ . \eqno(6)$$
The additional ${\bf C}$ term introduces an extra $U(1)$ symmetry,
the price one have to pay for having $SU(3)_{color}$ symmetry (no
appropriate Higgs fields). The symmetry is now $U(1)_{1}\times SU(2)
\times
U(1)_{2}\times U(3)$. To reduce it to the standard one, one have to
demand that the $U(1)_{1}$ part of the associated connection, Y, is
equal to the trace part of the $U(3)$ term and that the $U(1)_{2}$
part is equal to -Y [6]. A more elegant but equivalent treatment
can be found in [11]. This defines the algebraic structure of the
full standard model. The Dirac operator has the form:

$$D=\pmatrix{{\not \partial} \otimes Id & \gamma _{5}\otimes M^
{\dag}\cr \gamma _{5}\otimes M & {\not \partial}
\otimes Id\cr} \ , \eqno(7)$$
where

$$M=\pmatrix{0&0&0&0&0&0&0\cr 0&0&0&0&m_{e}&0&0\cr
0&0&0&0&0&m_{u}&0\cr 0&0&0&0&0&0&m_{d} \cr
0&m_{e}&0&0&0&0&0\cr 0&0&m_{u}&0&0&0&0\cr
0&0&0&m^{\dag}_{d}&0&0&0\cr} \ ,\eqno(8)$$
and the entries $m_{e}\ , m_{u}\ , m_{d}$ are positive definite
$N_{g}\times N_{g}$ matrices. The Yang-Mills functional is defined
by a representation $\pi :\Omega ^{*}(A)\rightarrow B(h)$ of the
differential algebra $\Omega ^{*}(A)$ in the Hilbert space h in
terms of bounded operators on h:

$$\pi \left( a_{0}da_{1}\dots da_{k}\right) = a_{0}i^{k}\left[
D,a_{1}\right] \dots \left[ D,a_{k}\right] \ . \eqno(9)$$
by

$${\it L_{YM}}=\frac{1}{4} Tr_{\omega}\left( \left( \pi ^{2}
\left( \theta \right) \right) D^{-4}\right) =\frac{1}{4} \int
d^{4}x Tr \left( tr \left( \pi ^{2}\left( \theta \right)
\right) \right) \ , \eqno(10)$$
where $\theta$ is the noncommutative curvature form, $\theta
=d\rho +\rho ^{2}$. $Tr_{\omega},\ Tr $ and $tr$ denote the
Diximier trace, trace over the matrices and trace over the
Clifford algebra, respectively [5, 6, 11, 12]. We have

$$\rho =\pmatrix{{\tilde A}_{1}\otimes Id &\gamma _{5}
\otimes H & 0 & 0 \cr
\gamma _{5} \otimes H^{\dag} & {\tilde A}_{2}\otimes Id & 0 & 0\cr
0 & 0& -{\tilde A}_{2} \otimes Id & 0\cr
0 & 0 & 0 & {\tilde A}_{colour}\otimes Id\cr}
\eqno(11)$$

$${\tilde A}_{1}=\pmatrix{ {\tilde A}^{3} & {\tilde A}^{1}
-i{\tilde A}^{2} \cr
{\tilde A}^{1} +i{\tilde A}^{2} & -{\tilde A}^{3} \cr}\eqno(12)$$

$${\tilde A}_{2}=i{\tilde A}^{0}=Tr{\tilde A}_{colour}  \eqno(13)$$
and $W^{+}= \frac{1}{\sqrt{2}}({\tilde A}^{1}-i{\tilde A}^{2}),
\ Z=\frac{1}{\sqrt{2}}({\tilde A}^{0}
+ {\tilde A}^{3})$ etc. The tilde sign has been used to denote
gauge fields of the
corresponding  algebras. After elimination of auxiliary fields and
Wick-rotating to Minkowski space we get
$$\begin{array}{ll}
{\it L_{YM}}= &\int \lbrace \frac{1}{4} N_{g}\left(
F_{\mu \nu}^{1}F^{1 \mu \nu}  + F_{\mu \nu}^{2}F^{2 \mu \nu} +
F_{\mu \nu}^{c}F^{c \mu \nu} \right) \cr
\ & + \frac{1}{2}Tr\left( MM^{\dag}\right) | \partial
H + A_{1}H - H^{\dag}A_{2} | ^{2}\cr
 \  & -\frac{1}{2} \left( Tr\left( MM^{\dag}\right) ^{2} - \left(
Tr MM^{\dag}\right) ^{2}\right) \left( HH^{\dag} -1\right)
^{2}\rbrace d^{4}x \cr \ .
\end{array}\eqno(14)$$
The fermionic action is given by

$$
\begin{array}{lll}
{\it L_{f}} & = & <\psi | D + \pi \left( \rho \right) | \psi >\cr
\ & = & \int \left( {\bar \psi } _{L} {\not D}\psi _{L} +
{\bar \psi } _{R} {\not D}\psi _{R} + {\bar \psi }_{L}H\otimes
M\psi _{R} + {\bar \psi }_{R}H^{\dag} \otimes M ^{\dag}
\psi _{L}\right) d^{4}x \end{array}\ , \eqno(15)$$
where we have included the $\pi (\rho)$ term into ${\not D}$.

\ \ \  Let us look closer at the full Lagrangian, ${\it L = L_{YM}
+ L_{f}}$. It has the standard form except for the $N_{g}$ factor
in front of the gauge field kinetic terms that comes from the
trace over generations. The analogous term in ${\it L_{f}}$ give
the sum over generations. We know that there are only three light
generations of fermions but is that all? We should count all
generations in ${\it L}$! This means that the coefficient in front
of the $F_{\mu \nu}F^{\mu \nu}$ terms should depend on $N_{g}$
and, in fact, give us information about the total numbers of
generations because it is absent from the fermionic part!
This is not true. The orthodox normalization is correct. We should
normalize the Diximier trace in (10) so that the coefficient
$N_{g}$ disappears. The simplest and natural solution is to
normalize $Tr$ so that $Tr Id_{N_{g}}=1$ [16]. This ensures also
that $Tr_{\omega}$ is always finite. There is a natural inner
product on the algebra of complex square matrices given by
$Tr(AB^{\dag})$. If one apply the Cauchy-Schwarz Inequality
to this inner product, one gets

$$Tr\left( MM^{\dag}\right) ^{2} \le \left( TrMM^{\dag}\right) ^{2}
\eqno(16)$$
 We cannot ensure the correct sign of the Higgs mass term without
the above normalization. The normalization of the trace $Tr$ leads
to

$$Tr\left( MM^{\dag}\right)^{2} \le N_{g} \left( TrMM^{\dag}
\right)^{2} \ . \eqno(17)$$
This means that for a big $N_{g}$ the coefficient $K=Tr\left(
MM^{\dag}\right) ^{2} - \left( TrMM^{\dag}\right) ^{2}$ may be
very large. In fact, it is possible that $K \rightarrow \infty$ if
the number of heavy  generations is unlimited. This force the
condition $HH^{\dag}=1$ in the Lagrangian and removes the Higgs
particle from the spectrum! If we are going to interpret the
Yukawa coupling in the standard way then we are not allowed to
arbitrary rescale the Higgs field and the limiting case leads to

$$m_{H}=\sqrt {2{{Tr\left( MM^{\dag}\right) ^{2} -
\left( TrMM^{\dag}\right) ^{2}} \over {TrMM^{\dag}}}}\  \
\rightarrow \infty \eqno(18)$$
as should be expected. The fermionic masses are in such a
(nonlinear) model by means of Yukawa couplings in a way analogous
to that of the standard model [13-15]. The fermionic part of the
Lagrangian given by Eq. (15) has the required form!

\ \ \ Another interesting possibility is to consider a
"more symmetric" version containing two $SU(2)$ factors. Then,
in order to have adjoint Higgs representations, one have to
extent the spacetime by two points for each $SU(2)$ factor and
identify the two copies [10, 16]. If the suggested above mechanism
realy works one gets a model that predicts several interesting
facts [14, 15, 17]. For example, the two- and three-vector-boson-
production process in $e^{+}e^{-}$ collisions at $\sqrt {s} =
500\  GeV$ (NLC) will give precise bounds for the parameters [18].

\ \ \ Let us conclude by saying that the BESS mechanism is a
necessary consequence of noncommutative version of the standard
model if there are many heavy generations. Such models are
discrete counterparts of the $CP^{n}$ sigma model obtained in
the Kaluza-Klein program [19] but far more realistic: they
predict interesting, experimentally verifiable facts. If we try
to preserve the standard interpretation of the mass scale of the
model [5, 6] then our case corresponds to infinitesimal distance
between copies of ordinary four-dimensional spacetime.

\ \ \ {\bf Acknowledgements}. The author would like to thank Prof.
R. K\" ogerler and dr K. Ko\l odziej for stimulating and helpful
discussions. This work has been supported in part by the Alexander
von Humboldt Foundation and the Polish Committee for Scientific
Research under the contract KBN-PB 2253/2/91.

\newpage
\subsection*{\ \ References}

\newcounter{bban}

\begin{list}
{[\arabic{bban}]}{\usecounter{bban}\setlength{\rightmargin}
{\leftmargin}}

\item J. S\l adkowski, Fortschr. Physik {\bf 38} (1990) 477.
\item R. Ma\'nka and J. S\l adkowski Phys. Lett. {\bf B224}
(1989) 97.
\item R. Ma\'nka and J. S\l adkowski Acta. Phys. Pol. {\bf B21}
(1990) 509.
\item J. S\l adkowski J. Phys. {\bf G16} (1990) L41.
\item A. Connes, in The interface of mathematics and physics
(Claredon, Oxford, 1990) eds . D. Quillen, G. Segal and S. Tsou.
\item A. Connes and J. Lott, Nucl. Phys. {\bf B} Proc. Suppl.
{\bf 18B} (1990) 29.
\item D. Kastler, Marseille preprints CPT-91/P.2910 and
CPT-91/P.2611 (1992).
\item R. Coquereau, G. Esposito-Far\'ese and G.Vaillant, Nucl.
Phys. {\bf B353} (1991) 689.
\item R. Coquereau, G. Esposito-Far\'ese and F. Scheck, INT. J. Mod.
Phys. {\bf A7} (1992) 6555.
\item A. H. Chamseddine, G. Felder and J. Fr\"ohlich, Phys. Lett.
{\bf B296} (1992) 109.
\item J. G. V\'arilly and J. M. Garcia-Bond\'ia, to be published in
J. Geom. Phys.
\item A. Connes, Publ. Math. IHES {\bf 62} (1983) 44.
\item R. Cosalbuoni, S. de Curtis, D. Dominici and R. Gatto,
Nucl. Phys. {\bf B282}, 235 (1987); Phys. Lett. {\bf B155},
(1985) 95.
\item G. Cvetic and R. K\"ogerler, Nucl. Phys. {\bf B328}
(1989) 342; {\it ibid.} {\bf B353}(1991) 462.
\item R. B\"onish and R. K\"ogerler, Int. J. Mod. Phys.
{\bf A7} (1992) 5475.
\item A. H. Chamseddine, G. Felder and J. Fr\"ohlich, ETH preprint,
ETH/TH/92/44 (1992).
\item A. H. Chamseddine, G. Felder and J. Fr\"ohlich, Nucl. Phys.
{\bf B395} (1993) 672.
\item R. B\"onish, C. Grosse-Knetter and R. K\"ogerler,
Bielefeld University preprint, BI-TP92/59 (1992).
\item J. H. Yoon, Mod. Phys. Lett. {\bf A7} (1992) 2611.

\end{list}

\end{document}